# Recent progress on electron- and magnon-mediated torques

Jia-Min Lai(来嘉敏)[1,2], Bingyue Bian(边冰玥)[1,2], Zhonghai Yu(于忠海)[1,2], Kaiwei Guo(郭凯卫)[1,2], Yajing Zhang(张雅静)[1,2], Pengnan Zhao(赵鹏楠)[1,2], Xiaoqian Zhang(张霄倩)[1,2], Chunyang Tang(汤春阳)[1,2], Jiasen Cao(曹家森)[1,2], Zhiyong Quan(全志勇)[1,2], Fei Wang(王飞)[1,2,†] and Xiaohong Xu(许小红)[1,2‡]

[1]*School of Chemistry and Materials Science, Key Laboratory of Magnetic Molecules and Magnetic Information Materials, Ministry of Education, Shanxi Normal University, Taiyuan 030031, China*

[2]*Research Institute of Materials Science, Shanxi Key Laboratory of Advanced Magnetic Materials and Devices, Shanxi Normal University, Taiyuan 030031, China*

The growing demand for artificial intelligence and complex computing has underscored the urgent need for advanced data storage technologies. Spin-orbit torque (SOT) has emerged as a leading candidate for high-speed, high-density magnetic random-access memory due to its ultrafast switching speed and low power consumption. This review systematically explores the generation and switching mechanisms of electron-mediated torques (including both conventional SOTs and orbital torques) and magnon-mediated torques. We discuss key materials that enable these effects: heavy metals, topological insulators, low-crystal-symmetry materials, non-collinear antiferromagnets, and altermagnets for conventional SOTs; $3d$, $4d$, and $5d$ transition metals for orbital torques; and antiferromagnetic insulator NiO- and multiferroic $BiFeO_3$-based sandwich structures for magnon torques. We emphasize that although key components of SOT devices have been demonstrated, numerous promising materials and critical questions regarding their underlying mechanisms remain to be explored. Therefore, this field represents a dynamic and rapidly evolving frontier in spintronics, offering significant potential for advancing next-generation information storage and computational technologies.

**Keywords:** spin-orbit torque, orbital torque, magnon torque, altermagnet

**PACS:** 75.70.Tj, 75.60.Jk, 75.76.+j, 72.25.-b

## 1. Introduction

The explosion of applications such as artificial intelligence and complex problem-solving has put forward higher requirements for data storage. Modern computer systems employ hierarchical data management, but the speed mismatch between slower storage units and the central processing unit creates a "storage wall", limiting computational power upgrades. Magnetic random-access memory (MRAM) offers fast read/write speeds comparable to cache memory while maintaining non-volatility like external storage. This makes it a promising solution to the "storage wall" issue through a non-von Neumann in-memory computing architecture[1-5]. Therefore, MRAM with high-speed, low-power, and non-volatile writing capabilities has become a core

---

objective in the development of future storage and integrated computing technologies.

MRAM has undergone three generations of development. The first-generation MRAM, known as Toggle-MRAM, manipulates the magnetization of a ferromagnetic layer (FM) through a current-induced Oersted field. However, this approach results in significant thermal losses, limiting device miniaturization and chip integration. Moreover, its magnetic tunnel junction utilizes an FM with in-plane magnetic anisotropy (IMA), which leads to edge magnetization and vortex effects as device fabrication processes shrink, ultimately reducing reliability. The second-generation MRAM, based on spin-transfer torque (STT), represents a major advancement toward commercialization[6-9]. Nevertheless, STT-MRAM faces several challenges, including non-separated read/write paths that can introduce read/write disturbances, tunneling currents that increase the risk of barrier breakdown, and incubation delays that limit switching speed[5, 10-12]. In contrast, third-generation spin-orbit torque (SOT)-MRAM leverages spin-orbit coupling (SOC) in a non-magnetic (NM) layer to convert charge currents into spin currents[13-16]. These spin-polarized electrons accumulate at the FM/NM interface, exerting torques on the FM and inducing magnetization switching. SOT-MRAM offers distinct advantages, including separate read/write paths, reduced read currents to mitigate barrier degradation, and sub-nanosecond switching speeds[4, 17-22]. These attributes establish SOT-MRAM as a leading candidate for high-speed cache memory in the post-Moore era.

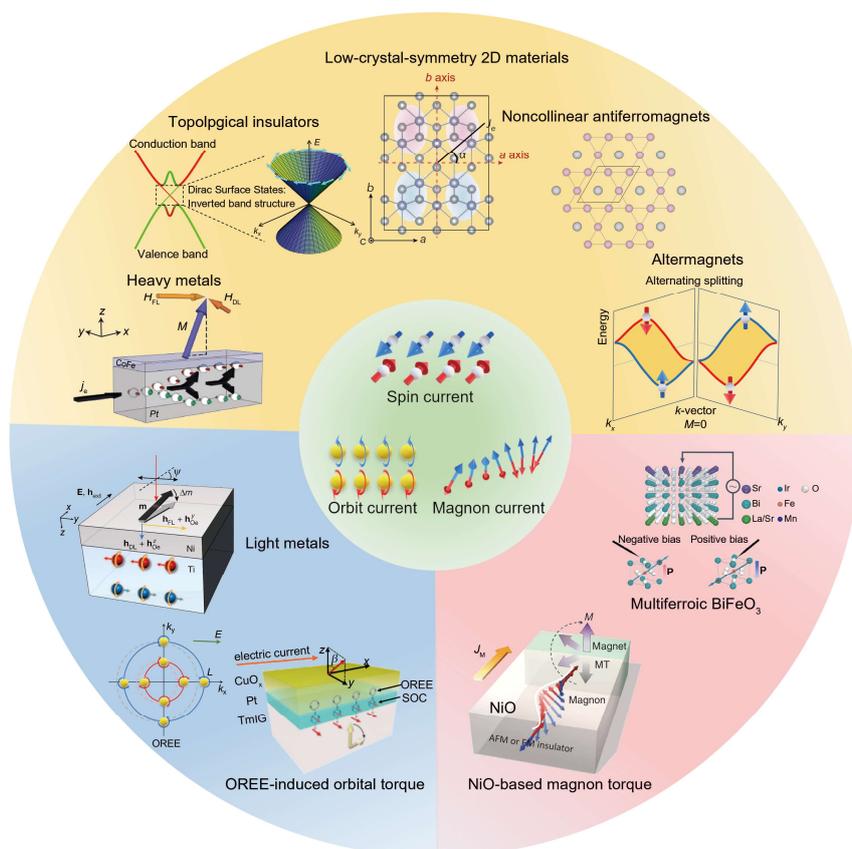

**Fig. 1.** Schematic overview of materials and device structures for conventional spin-orbit torques, orbital torques, and magnon torques. Reprinted with permission from Ref. [23-28]. Copyright 2013 by Springer Nature. Copyright 2019 by American Association for the Advancement of Science.

Since the concept of SOT was demonstrated for manipulating FMs with IMA[29], the SOT-MRAM with in-plane MTJ has been demonstrated to be compatible with back-end-of-line processes[30, 31]. The collinear relationship between the spin polarization and magnetization directions in SOT enables the field-free manipulation of FM with IMA. However, further miniaturization limits the scalability potential of in-plane SOT-MRAM. The significant breakthrough of SOT-driven switching of a perpendicular magnetization was achieved in Pt/Co systems[13, 32]. The spin current generated in the Pt layer exerts torques on the Co layer with PMA, enabling its magnetization switching with the assistance of an external magnetic field. The perpendicular switching of FM marks a milestone in the development of SOT-based memory and logic devices. Subsequently, systematic investigations into spin-source materials are conducted, initially focusing on heavy metals[13, 14, 23, 33] and topological insulators (TIs)[16, 34-41] due to their high charge-to-spin conversion efficiency. These materials are primarily explored to reduce the critical switching current density required for magnetization switching. However, researchers soon recognized that practical SOT device applications require not only low power consumption but also field-free switching of perpendicular magnetization[42]. This has led to remarkable interest in exploring symmetry-breaking device geometries with gradient electric field[43, 44], composition[45-47], or thickness[47, 48], and $z$-spin generating materials[42], e.g., low-crystal-symmetry materials[27, 49-60], non-collinear antiferromagnets (NCAFMs)[61-74], altermagnets[73-77], and spin Hall metals[78].

Beyond spin, the orbital degree of freedom in electrons has also been utilized to manipulate FMs. Compared to spin angular momentum, orbital angular momentum exhibits a longer diffusion length and higher orbital Hall conductivity (OHC)[26, 79-81]. Moreover, strong orbital currents have been observed in cost-effective 3$d$ and 4$d$ light metals[26, 80, 82-87], enhancing the industrial compatibility of orbital torque (OT) devices. OT has emerged as a promising alternative to and potential advancement over conventional SOT, attracting significant research interest in spintronic applications. Additionally, magnon currents, as spin-wave excitations of magnetic moments, can convey spin angular momentum in insulating magnetic materials while exhibiting long spin diffusion lengths[88-90] and ultrafast transport speeds[91]. These unique properties make magnon torque (MT) a promising avenue for developing next-generation ultrafast and low-power magnetic storage devices. In this review, we systematically introduce the generation and switching mechanisms of conventional SOT, OT, and MT, followed by a discussion on recent material innovations and device advancements in these three fields, as summarized in Fig. 1. Finally, we provide a comprehensive summary and offer an outlook on future developments in SOT-based spintronic devices.

## 2. Conventional spin-orbit torque
## 2.1. Spin Hall effect and Rashba-Edelstein effect

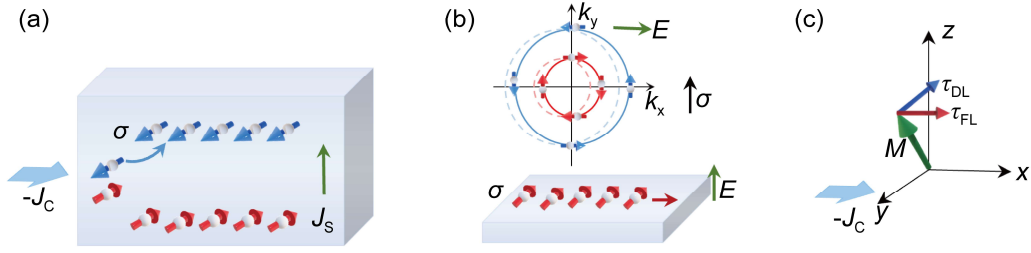

**Fig. 2.** The principal diagram of (a) spin Hall effect and (b) Rashba-Edelstein effect. (c) Direction of anti-damping torque (blue arrow) and field-like torque (red arrow). Green arrows indicate the magnetization direction of the ferromagnetic layer.

Spin current generation is primarily governed by two well-established mechanisms: the bulk spin Hall effect (SHE) and the interfacial Rashba-Edelstein effect (REE). The SHE occurs in an NM layer with strong SOC, where spin Berry curvature plays a crucial role. When an electric field is applied, the resulting charge current ($J_C$) is converted into a transversely polarized spin current ($J_S$)[92, 93], leading to spin accumulation at the NM/FM interface. The direction of spin polarization ($\sigma$) is orthogonal to both $J_C$ and $J_S$, as illustrated in Fig. 2(a). The charge-to-spin conversion process can be quantitatively described by the relation: $J_S = (\hbar/2e)\theta_{SH}(J_C \times \sigma)$, where $\hbar$ and $e$ are the reduced Planck constant and the elementary charge, respectively[10]. $\theta_{SH}$ represents the spin Hall angle, a key parameter characterizing the efficiency of charge-to-spin conversion and serving as a crucial indicator of SOT performance. The power consumption ($W$) required for magnetization switching via SOT is proportional to $J_C^2/\sigma_j$, where $\sigma_j$ denotes the charge conductivity[11]. This relationship can be further expressed as: $W \propto 1/(\theta_{SH}^2 \cdot \sigma_j) = 1/(\theta_{SH} \cdot \sigma_{SH})$, where $\sigma_{SH} = \theta_{SH} \cdot \sigma_j$ defines the spin Hall conductivity (SHC). To achieve low-power SOT device, the spin source material must exhibit both a high $\theta_{SH}$ and a large $\sigma_j$, thereby maximizing $\theta_{SH} \cdot \sigma_{SH}$. In contrast, the REE arises from interfacial SOC in systems with broken inversion symmetry[94, 95]. When a perpendicular built-in electric field is present at the interface, electrons moving along the x-axis experience an effective magnetic field along the y-axis, leading to spin accumulation at the interface, as illustrated in Fig. 2(b). This interfacial effect enables efficient spin generation and manipulation, complementing the bulk-dominated SHE mechanism.

In NM/FM heterostructures, the coexistence of the SHE and the REE gives rise to ongoing debates regarding the precise origin of the generated spin current. However, it is widely recognized that SOT can exert two distinct torque components on the magnetization ($M$) by $\sigma$[96-98], as shown in Fig. 2(c). The anti-damping torque ($\tau_{DL} \sim M \times (\sigma \times M)$) drives $M$ toward alignment with $\sigma$, facilitating magnetization switching. Meanwhile, the field-like torque ($\tau_{FL} \sim (M \times \sigma)$) induces magnetic moment precession around the direction of $\sigma$. For an FM with perpendicular magnetic anisotropy (PMA), the SOT-induced torque tilts the magnetization towards the y-direction. Upon removal of the current, thermal fluctuations cause the magnetization to stochastically switch to either the +z or −z direction in the absence of an external magnetic field. Achieving deterministic switching of perpendicular magnetization necessitates an additional in-plane magnetic field along the current direction to break mirror symmetry. However, this reliance on an external magnetic field poses a significant challenge for high-density device integration. To overcome this limitation, research efforts have not only focused on optimizing heavy metals and TIs to enhance the spin Hall angle but have also shifted toward eliminating the need for an external field and achieving field-free switching of perpendicular magnetization. Consequently, novel structural designs and material innovations, such as low-crystal-symmetry materials[27, 49-60], NCAFMs[61-72], and altermagnets[73-77], have been extensively explored. In the following sections, we review the progress in conventional SOT systems from the perspective of material development.

## 2.2. Heavy metals and topological insulators

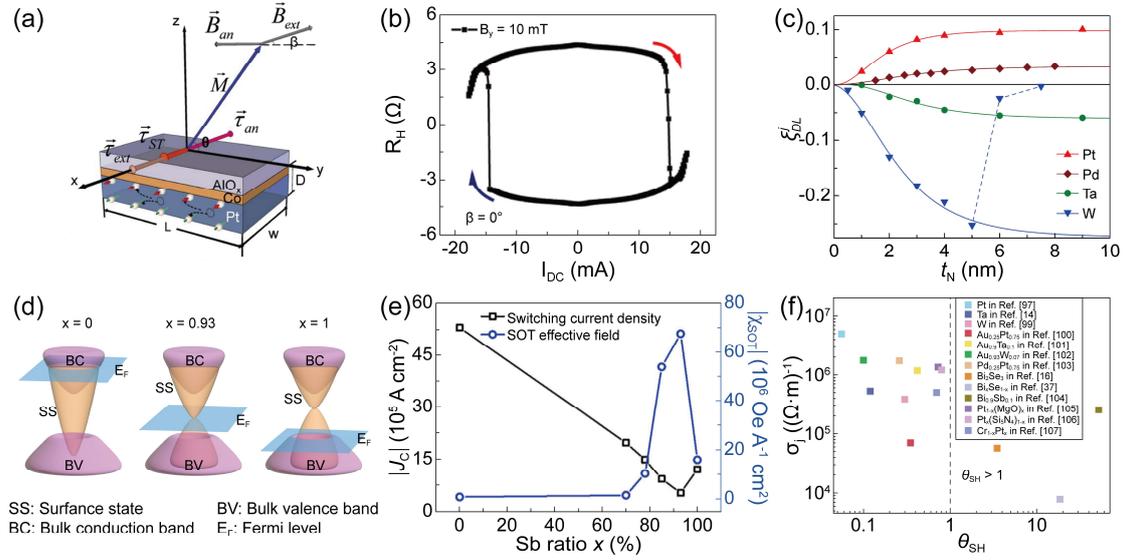

**Fig. 3.** Spin-orbit torque based on heavy metals and topological insulators. (a) Illustration of spin-orbit torque in a Co/Pt/AlO$_x$ heterostructure. (b) Current-induced magnetization switching in Pt/Co/AlO$_x$ heterostructures. Reprinted with permission from Ref. [15]. Copyright 2012 by American Physical Society. (c) Anti-damping torque efficiency as a function of thickness in heavy metal/Co/AlO$_x$ layers, where heavy metals are Pt, Pd, Ta, and W. Reprinted with permission from Ref. [98]. Copyright 2019 by American Physical Society. (d) Schematic of the Fermi level positions for different Sb ratios ($x = 0, 0.93, 1$) of (Bi$_{1-x}$Sb$_x$)$_2$Te$_3$. (e) Switching current density ($J_C$) and spin-orbit torque-induced effective field ($\chi_{SOT}$) as a function of Sb ratios ($x$) in (Bi$_{1-x}$Sb$_x$)$_2$Te$_3$. Reprinted with permission from Ref. [38]. Copyright 2022 by American Physical Society. (f) The charge conductivity ($\sigma_j$) and spin Hall angle ($\theta_{SH}$) for various materials[14, 16, 37, 97, 99-107].

Early investigations into spin source materials primarily focused on heavy metals and their alloys, such as Pt[13, 15, 23, 97, 108, 109], Ta[14, 23], W[99], Au$_{0.25}$Pt$_{0.75}$[100], Pd$_{0.25}$Pt$_{0.75}$[103], Pt$_{1-x}$(MgO)$_x$[105], Pt$_x$(Si$_5$N$_4$)$_{1-x}$[106], Cr$_{1-x}$Pt$_x$[107]. The strong SOC in these materials enables efficient charge-to-spin conversion, making them key candidates for SOT applications. Miron et al. attributed the current-induced perpendicular magnetization switching in Pt/Co heterostructures to REE mechanism[13]. However, Liu et al. later demonstrated that the spin current originates from the SHE in Pt[15], as illustrated in Fig. 3(a). The positive spin Hall angle of Pt results in a clockwise SOT switching loop under a positive external magnetic field, as shown in Fig. 3(b).

Subsequent studies have explored the fundamental origins of spin currents in various heavy-metal-based systems[98], as summarized in Fig. 3(c). The prevailing consensus is that in most heavy metals, SOT is primarily governed by the SHE. However, the presence and contribution of the REE remain dependent on specific interface conditions[10]. Even within the same material system, variations in fabrication techniques across different research groups can lead to substantial differences in interfacial properties, which in turn significantly influence the magnitude and role of the REE.

In the pursuit of higher charge-spin conversion efficiency, TIs have emerged as highly promising materials due to their spin-momentum-locked topological surface states (TSSs)[16, 39, 110, 111]. When a charge current flows through the TSS, the resulting spin polarization is inherently locked perpendicular to the momentum direction. This unique property enables TIs to achieve exceptionally high charge-spin conversion efficiencies, even exceeding 1[16, 37, 104]. As a result, the critical switching current density ($J_C$) required for perpendicular magnetization switching is remarkably reduced to $\sim 10^5$-$10^6$ A·cm$^{-2}$[37, 38, 40, 41, 104, 112-114], which is one to two orders of magnitude lower than that in heavy metal-based systems[13, 15]. The initial experimental demonstration of SOT based on TIs was reported in (Bi$_{0.5}$Sb$_{0.5}$)$_2$Te$_3$/(Cr$_{0.08}$Bi$_{0.54}$Sb$_{0.38}$)$_2$Te$_3$ bilayers[34] and Bi$_2$Se$_3$/Ni$_{81}$Fe$_{19}$ (permalloy, Py) heterostructures[16], respectively. Since then, extensive research has been conducted on TI-based SOT systems such as Bi$_2$Se$_3$[35, 36, 112, 115, 116], Bi$_2$Te$_3$[40], Bi$_x$Sb$_{1-x}$[37, 104], and (BiSb)$_2$Te$_3$[38, 41, 112-114, 117-119]. A key advantage of TI-based SOTs is the tunability of SOT efficiency through modulation of the Fermi level $E_F$[38, 117, 120]. Wu et al. demonstrated that by tuning the Sb composition $x$ in (Bi$_{1-x}$Sb$_x$)$_2$Te$_3$, $E_F$ can be shifted from the bulk conduction band ($x = 0$) to the topological surface band ($x = 0.93$), and further into the bulk valence band ($x = 1.0$)[38], as illustrated in Fig. 3(d). Notably, the SOT efficiency reaches its peak near the Dirac point ($x = 0.93$), where bulk conduction is minimized while surface-state conductivity is maximized, as shown in Fig. 3(e). Beyond doping[34, 36, 38, 41, 114, 119, 120], TSSs and their associated SOT efficiencies can be modulated through various approaches, including thickness control[36], electric field application[117], interface engineering[115], ultrafast optical pumping[121], and strain engineering[122]. These methods provide additional avenues to tailor spin polarization and optimize SOT performance, further advancing the potential of TIs for next-generation spintronic applications.

For the realization of low-power SOT devices, an ideal spin source material should not only exhibit a high $\theta_{SH}$ but also possess high $\sigma_j$. If the spin source material has low $\sigma_j$, a significant portion of the current will pass through the adjacent FM, leading to excessive power dissipation and reduced device efficiency. Figure 3(f) illustrates the inherent trade-off between $\theta_{SH}$ and $\sigma_j$ across various materials. For instance, Pt offers the highest $\sigma_j$ but suffers from a relatively low $\theta_{SH}$[97], whereas TIs exhibit a large $\theta_{SH}$ but are constrained by poor $\sigma_j$[16, 37, 104]. Addressing this trade-off remains a critical challenge in the development of high-efficiency SOT devices. Moreover, most conventional spin source materials exhibit high crystal symmetry, which allows for the generation of only $y$-polarized spin currents, necessitating an external magnetic field along the current direction to achieve deterministic switching of perpendicular magnetization. To eliminate the requirement for assistive magnetic fields, researchers have explored various approaches to break system symmetry, including structural asymmetry[48, 123], gradient electric field[43, 44], exchange interaction engineering[124, 125], magnetization tilting[126-129], and hybrid manipulation techniques[43, 130]. Besides, the interlayer Dzyaloshinskii–Moriya interaction usually realized in Pt/Co stacks[131, 132] provides a field-free switching strategy that is compatible with perpendicular SOT-MRAM. While these methods can enable field-free magnetization switching, they introduce additional complexities in device design and may lead to extrinsic effects. A promising alternative is leveraging materials with inherently low symmetry to generate $z$-polarized spin currents, thereby breaking system symmetry[42].

## 2.3. Low-crystal-symmetry materials

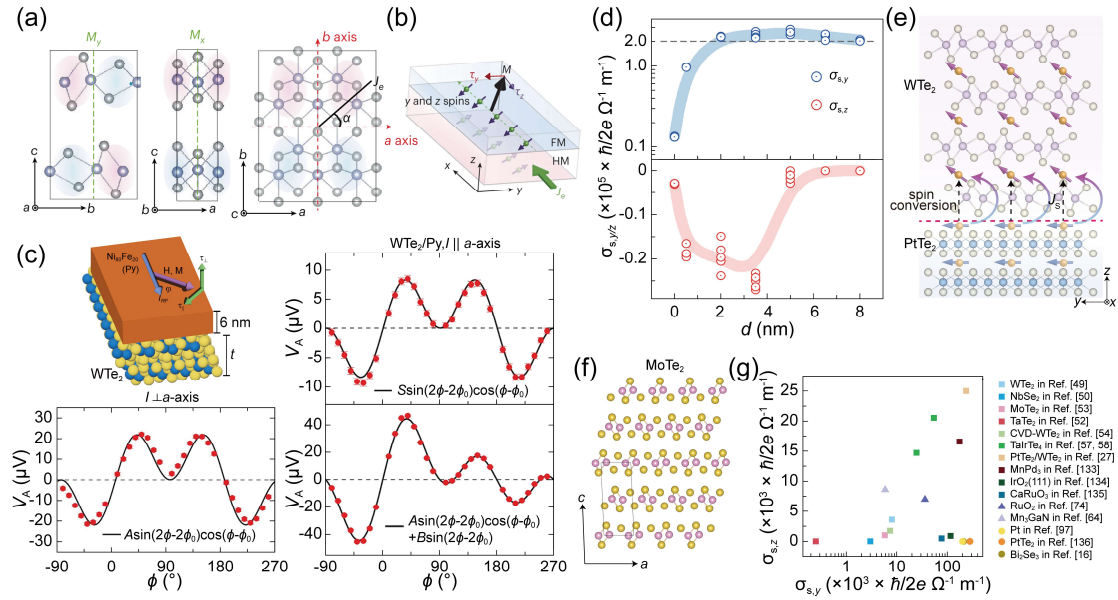

**Fig. 4.** Spin-orbit torque based on low-crystal-symmetry materials. (a) Crystal structure of $T_d$-WTe$_2$ ($Pmn2_1$), where $M_x$ and $M_y$ represent the pure mirror and glide mirror plane, respectively. (b) Schematic of the spin-orbit torque device in a heavy metal/ferromagnet bilayer, where the in-plane charge current can generate both $y$- and $z$-polarized spins. The $z$-polarized spins exert an out-of-plane anti-damping torque, which enables field-free magnetization switching of the ferromagnetic layer. Reprinted with permission from Ref. [27]. Copyright 2024 by Springer Nature. (c) Schematic of WTe$_2$/Py bilayers and the angular dependence of spin-torque ferromagnetic resonance signals. $\varphi$ is the angle between the current direction and the external magnetic field. Reprinted with permission from Ref. [49]. Copyright 2016 by Springer Nature. (d) The $d$ dependence of in-plane and out-of-plane spin Hall conductivity ($\sigma_{s,y}$ and $\sigma_{s,z}$) in PtTe$_2$ ($d$)/WTe$_2$ (8–$d$)/Py heterostructures. Reprinted with permission from Ref. [27]. Copyright 2024 by Springer Nature. (e) Spin-to-spin conversion from an in-plane polarized spin Hall current generated in PtTe$_2$ layer to an out-of-plane polarized spin current in WTe$_2$ layer. (f) $a$-$c$ plane of the $\beta$-MoTe$_2$ crystal structure, where the mirror plane is within the page and the Mo chains run into the page. (g) The in-plane and out-of-plane spin Hall conductivity for various materials[16, 27, 49, 50, 52-54, 57, 58, 64, 74, 97, 133-136].

Breaking lateral symmetry at the film surface has emerged as a common strategy for generating $z$-spins. This approach requires the absence of any $n$-fold ($n > 1$) rotational symmetry about the surface normal (z-axis) while allowing at most a single mirror plane to be preserved in the lattice structure. Such symmetry breaking can be achieved in specific crystallographic planes of low-crystal-symmetry materials. The first reported low-crystal-symmetry material capable of generating z-spins is the Weyl semimetal $T_d$-phase $WTe_2$[49], whose crystal structure is depicted in Fig. 4(a). $WTe_2$ belongs to the $Pmn2_1$ space group, where glide symmetry is broken along the low-symmetry a-axis, while mirror symmetry along the b-axis is preserved. When a charge current is applied along the a-axis, both y- and z-polarized spin currents are generated. These spin currents give rise to in-plane ($\tau_y$) and out-of-plane ($\tau_z$) anti-damping torques, as well as field-like torques on the adjacent FM[27], as illustrated in Fig. 4(b). The first experimental demonstration of the out-of-plane anti-damping torque in $WTe_2$/Py bilayers is reported by MacNeill et al. using spin-torque ferromagnetic resonance (ST-FMR) measurements[49, 137]. The mixed voltage signal ($V_{mix}$) obtained from ST-FMR can be decomposed into symmetric ($V_S$) and antisymmetric ($V_A$) components, which correspond to the in-plane ($\tau_\parallel$) and out-of-plane ($\tau_\perp$) torques, respectively. Figure 4(c) presents the angular ($\phi$) dependence of $V_S$ and $V_A$[49]. When the current is applied along the a-axis, the curves can be fitted using $V_S = A\sin2\phi\cos\phi$ and $V_A = B\sin2\phi\cos\phi + C\sin2\phi$, confirming the presence of an out-of-plane anti-damping torque and the corresponding generation of z-spins. In contrast, when the current flows along the b-axis, the fitting follows $V_A = B\sin2\phi\cos\phi$, which is consistent with the absence of an out-of-plane anti-damping torque. The ability to generate z-spins has enabled field-free switching of perpendicular magnetization in various $WTe_2$/FM heterostructures. Experimental demonstrations have been reported across a diverse range of FMs, including two-dimensional (2D) magnets[56, 59, 138, 139], perovskite oxides[140], and metallic multilayers[141, 142].

Despite its ability to generate z-spins, $WTe_2$ suffers from a critical drawback: low y-SHC ($\sigma_{s,y} = 8 \times 10^3$ ($\hbar/2e$) ($\Omega \cdot m$)$^{-1}$), which leads to huge current shunting through the FM, thereby increasing device power consumption. To address this limitation, Wang et al. fabricated a $PtTe_2$/$WTe_2$ heterostructure, leveraging the high SHC of $PtTe_2$[27]. Their study demonstrates that the incorporation of $PtTe_2$ not only enhances the y-SHC ($\sigma_{s,y} = 2.32 \times 10^5$ ($\hbar/2e$) ($\Omega \cdot m$)$^{-1}$), but also significantly improves the z-SHC ($\sigma_{s,z} = 0.25 \times 10^5$ ($\hbar/2e$) ($\Omega \cdot m$)$^{-1}$), as shown in Fig. 4(d). This enhancement in z-SHC is attributed to the spin-to-spin conversion (Fig. 4(e)), wherein y-spins generated in $PtTe_2$ are converted into z-spins in $WTe_2$. When used for perpendicular magnetization switching, the $PtTe_2$/$WTe_2$ heterostructure exhibits a 33-fold reduction in power consumption compared to $WTe_2$ alone[27].

The $1T'$-$MoTe_2$ monolayer is isostructural to the $WTe_2$ monolayer but retains the inversion symmetry in bulk crystal and breaks the lateral symmetry at the surface (Fig. 4(f))[53, 143]. In $MoTe_2$/Py, the z-spins are detected when $MoTe_2$ is in the monolayer and trilayer forms, but diminish to zero in the bilayers form, which demonstrates that the breaking of lateral surface symmetry rather than bulk inversion symmetry, is the necessary condition for producing z-spins[53]. The ternary Weyl semimetal $TaIrTe_4$ has a crystal structure similar to that of $WTe_2$, where Ta and Ir orderly replace W, thereby preserving its intrinsically low crystal symmetry. However, it exhibits a higher SHC due to its band structure and Fermi surface. Experiments have demonstrated that TaIrTe$_4$ can provide $\sigma_{S,y}$ of $5.437\times10^5$ ($\hbar/2e$) ($\Omega\cdot$m)$^{-1}$ and $\sigma_{S,z}$ of $2.065\times10^4$ ($\hbar/2e$) ($\Omega\cdot$m)$^{-1}$[57, 58, 60], which are one order of magnitude higher than those of WTe$_2$ and MoTe$_2$. The $z$-spins have also been experimentally observed in non-2D materials, including MnPd$_3$[133], IrO$_2$[134, 144], and CaRuO$_3$[135] (Fig. 4(g)), where epitaxial strain engineering is utilized to induce specific lattice orientations that break lateral symmetry. This structural design facilitates the generation of $z$-polarized spin currents, further expanding the material landscape for field-free switching applications.

## 2.4. Non-collinear antiferromagnets

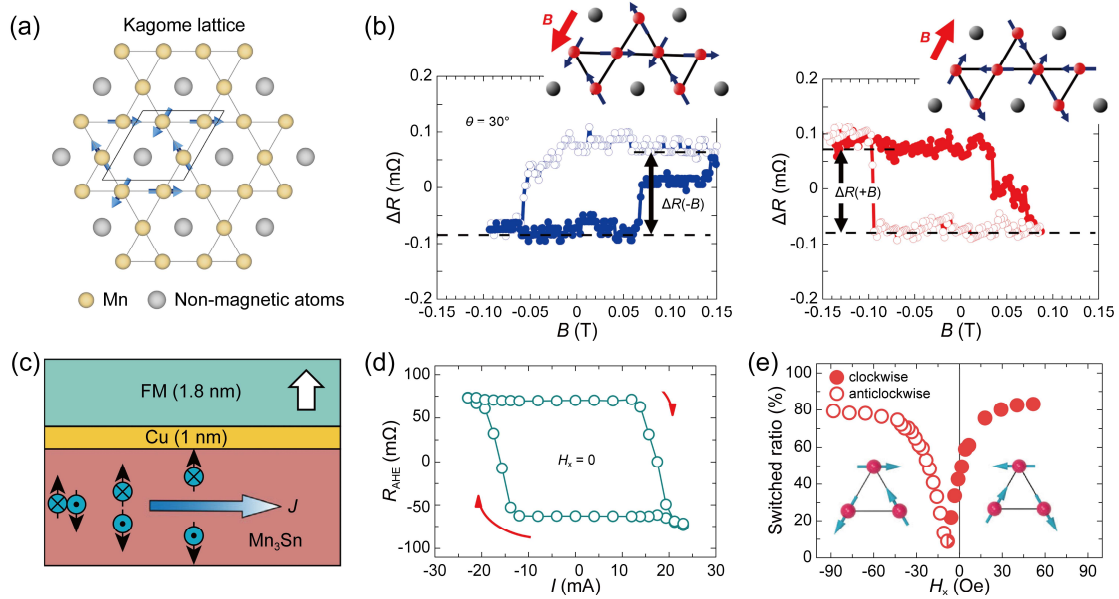

**Fig. 5.** The current-driven magnetization switching based on non-collinear antiferromagnets. (a) Kagome arrangements of Mn-based compounds, leading to a triangular magnetic order. (b) Spin accumulation at the Mn$_3$Sn-NiFe interface: magnetic-field dependence of resistance measured between the NiFe and Cu electrodes on the surface of Mn$_3$Sn single crystal. The results are obtained after saturation of Mn$_3$Sn with a magnetic field ($B$) of $-0.75$ T (blue) and $+0.75$ T (red), respectively. Insets show the corresponding spin structures. Reprinted with permission from Ref. [62]. Copyright 2019 by Springer Nature. (c) The schematic of Mn$_3$Sn/Cu/ferromagnet structure. (d) Current-induced perpendicular magnetization switching in the absence of magnetic field. (e) The switching polarity and switching ratio under different magnetic fields. The insets illustrate two opposite Mn$_3$Sn domains to generate out-of-plane spin currents, resulting in the clockwise and anticlockwise switching polarities. Reprinted with permission from Ref. [67]. Copyright 2022 by Springer Nature.

In addition to low-crystal-symmetry materials, low-magnetic-symmetry materials can also generate $z$-spins, enabling field-free switching of perpendicular magnetization. These materials include NCAFMs[64-72, 145] and altermagnets[76, 77] (discussed in section 2.5). Unlike conventional antiferromagnets (AFMs), low-magnetic-symmetry materials break $\hat{P}\hat{T}$ and $\hat{T}\hat{t}$ symmetries, where $\hat{P}$ and $\hat{T}$ denote space inversion and time reversal, respectively, and $\hat{t}$ indicates a half-unit-cell translation[146]. Consequently, their band structures exhibit spin splitting, allowing for the generation of spin currents along specific crystallographic directions, even in the absence of strong SOC. Moreover, NCAFMs and altermagnets combine the advantages of both FMs and AFMs, offering efficient spin current generation and detection, non-volatility, negligible stray fields, and ultrafast spin dynamics, while overcoming the disadvantages of these conventional systems[72, 147, 148]. As a result, NCAFMs and altermagnets have garnered significant interest for their potential applications in next-generation memory and computing technologies.

In spintronics, research on NCAFMs has predominantly focused on Mn-based compounds, where Mn atoms form a distinctive kagome lattice (Fig. 5(a)). In these materials, the magnetic moments of Mn atoms are oriented at 120º relative to one another, giving rise to a triangular chiral spin texture[72]. The Mn triangles are interconnected by non-magnetic atoms, typically transition or post-transition metal elements. A key feature of NCAFMs is their kagome spin texture, which breaks time-reversal symmetry and lifts the spin degeneracy of the Fermi surface, leading to the emergence of the anomalous Hall effect (AHE)[149] and the magnetic spin Hall effect (MSHE)[62, 67, 69, 150, 151]. Unlike the conventional SHE, the spin polarization generated by the MSHE is not necessarily perpendicular to both the charge current and spin current directions. Additionally, the spin polarization is highly dependent on the magnetic domain state of NCAFMs[62], allowing for active control over both the magnitude and direction of spin currents. A significant experimental breakthrough in this field is achieved by Yoshichika Otani's group[62], demonstrating a sign reversal in the hysteretic resistance upon switching a pre-aligned magnetic field in $Mn_3Sn$/NiFe heterostructures, as shown in Fig. 5(b). This provides direct evidence that the MSHE in $Mn_3Sn$ exhibits odd symmetry under time-reversal operations. The same research group subsequently provides further direct observations of $z$-spin generation via the MSHE, revealing that $z$-spins in $Mn_3Sn$ exert a magnetic spin Hall torque on an adjacent FM[151]. Moreover, they demonstrate that the magnitude of this torque can be tuned by adjusting the orientation of the magnetic octupoles. Following these pioneering studies, MSHE has also been observed in other Mn-based material systems[64, 66, 69-71, 152, 153], further expanding its potential applications in spintronic devices.

Field-free switching of perpendicular magnetization has been experimentally demonstrated in various NCAFMs-based systems, including $Mn_3Sn$[67, 68], $Mn_3Pt$[69], $Mn_3Ga$[70], $Mn_3Ir$[71], and $Mn_3SnN$[66]. Hu et al. fabricated $Mn_3Sn$/Cu/Co/Ni heterostructures (Fig. 5(c)) and successfully achieved field-free switching of perpendicular magnetization (Fig. 5(d))[67]. The switching ratio, defined as the change in transverse resistance during switching divided by the anomalous Hall resistance, is found to be approximately 60%. Notably, both the switching ratio and polarity of the FM could be actively modulated by reorienting the magnetic domains of $Mn_3Sn$ (Fig. 5(e))[67]. This discovery underscores the critical role of domain control in tailoring spin-torque effects in NCAFMs. Following this work, domain-controlled perpendicular magnetization switching has been observed in other heterostructures where NCAFMs serve as the spin source[66, 69-71].

## 2.5. Altermagnets

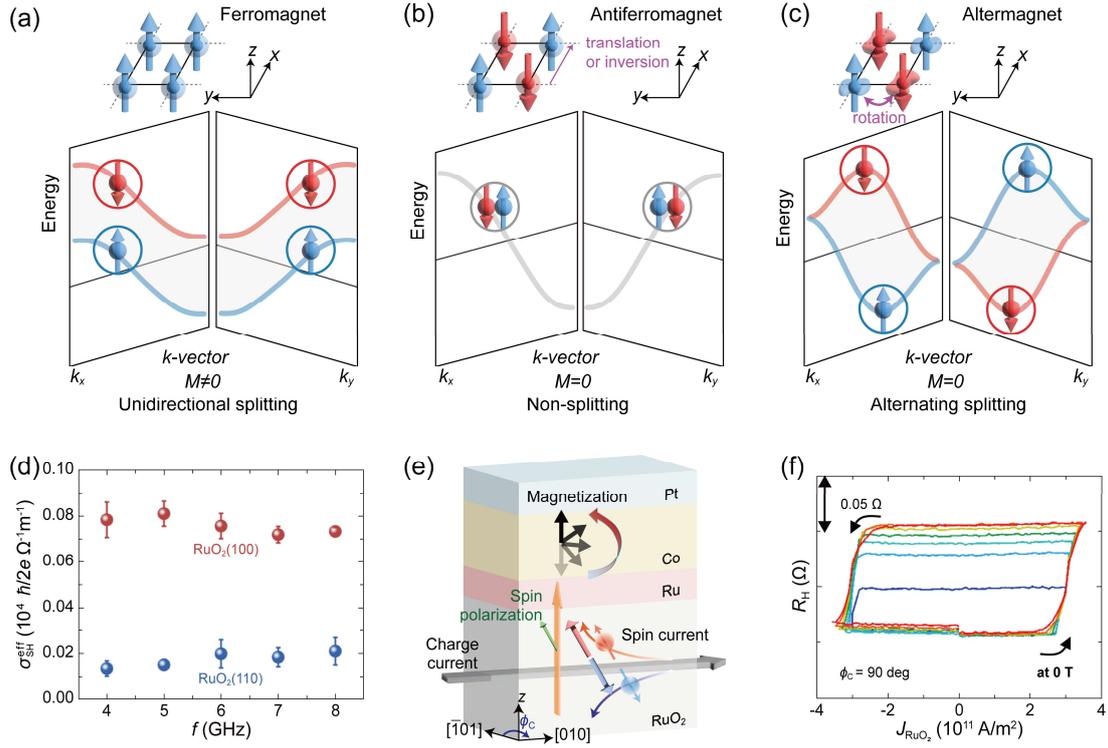

**Fig. 6.** The spin-split effect and field-free switching of perpendicular magnetization based on altermagnets. Schematic diagrams of spin configurations and electronic bands of (a) collinear ferromagnet, (b) antiferromagnet, and (c) altermagnet. (d) The calculated spin torque conductivity ($\sigma_{SH}^{eff}$) at different microwave frequencies. Reproduced with permission from Ref. [73]. Copyright 2022, The Authors, Published by American Physical Society. (e) Schematic illustration of $RuO_2(101)/Ru/Co/Pt$ multilayer. $\phi_C$ represents the angle between the direction of charge current and the $[\bar{1}01]$ direction of $RuO_2$. (f) Current-induced perpendicular magnetization switching under $\phi_C = 90°$. Reprinted with permission from Ref. [76]. Copyright 2022 by American Physical Society.

Altermagnets represent a distinct class of magnetic materials, positioned alongside FMs and AFMs[147, 148, 154] (Figs. 6(a)-6(c)). Unlike conventional FMs, which exhibit nonzero net magnetization and spin-split electronic bands, and AFMs, which feature zero net magnetization and spin-degenerate electronic bands, altermagnets possess a unique combination of both characteristics. Specifically, altermagnets maintain zero net magnetization while exhibiting spin-split electronic bands, which endow them with several advantageous properties: zero stray fields, ultrafast spin dynamics, and spin current generation[147, 148, 155]. The origin of these properties lies in the unique magnetic symmetry of altermagnets, where opposite-spin sublattices are interconnected through crystal-rotation symmetries (proper or improper, symmorphic or non-symmorphic) rather than through translation or inversion symmetries, as seen in conventional AFMs. A growing number of materials have been experimentally identified as altermagnets, including $RuO_2$[156, 157], MnTe[158-162], CrSb[163-168], $Mn_5Si_3$[169], $KV_2Se_2O$[170], and $RbV_2Te_2O$[171]. However, it is important to note that while $MnTe_2$ exhibits spin-split electronic bands[172], it does not strictly belong to the altermagnet family due to its non-coplanar antiferromagnetic ordering.

RuO$_2$ is a *d*-wave altermagnet characterized by two spin-degenerate nodal surfaces. It is currently the only known altermagnet capable of generating spin currents through the spin-split effect (SSE). It crystallizes in a tetragonal structure belonging to the *P*4$_2$/*mnm* space group. However, due to the inequivalent local crystal environments of its two opposite spin sublattices, RuO$_2$ exhibits momentum-space spin-split electronic bands, making it an efficient spin current generator[75]. The spin polarization direction of the spin current generated in RuO$_2$ is aligned with its Néel vector along the [001] axis[73-75]. As a result, the spin current direction can be precisely controlled by tuning the crystal orientation of RuO$_2$. For instance, in (100)-oriented RuO$_2$ films, a charge current induces an additional time-reversal-odd spin current through the SSE[73]. However, in (110)-oriented RuO$_2$ films, a charge current generates only *y*-polarized spin currents via the SHE[73]. This configuration reduces the charge-to-spin conversion efficiency compared to (100)-oriented RuO$_2$ films, as the contribution from the SSE is absent (Fig. 6(d)).

To generate *z*-spins via the SSE in RuO$_2$, the Néel vector of RuO$_2$ must deviate from, but not be perpendicular to, the film surface. Bose et al. experimentally demonstrated *z*-polarized spin current generation by fabricating a (101)-oriented RuO$_2$ film[74]. When a current was applied along the [010] axis, they observed a tilted spin current forming an angle of approximately 35º relative to the out-of-plane direction[74]. Karube et al. then designed a RuO$_2$ (101)/Ru/Co/Pt multilayer structure, where the Co layer exhibits PMA[76]. By applying a current along the [010] axis, they successfully generated a tilted spin current (including *x*-, *y*-, and *z*-spins), enabling field-free switching of the Co layer[76] (Figs. 6(e) and 6(f)). The field-free switching of perpendicular magnetization based on RuO$_2$ has been demonstrated by several independent groups[173-176].

## 3. Orbital torque
### 3.1. Mechanism of orbital torque

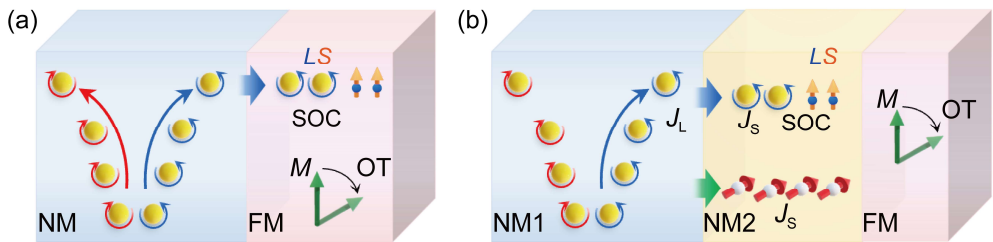

**Fig. 7.** The illustration of magnetization switching by orbital torque based on the orbital Hall effect in (a) non-magnetic layer/ferromagnetic layer (NM/FM) heterostructure and (b) NM1/NM2/FM heterostructure, where NM2 is inserted as an orbital-spin conversion layer.

Electrons possess not only spin angular momentum but also orbital angular momentum. Over the past few decades, orbital angular momentum has been conventionally assumed to be quenched by the crystal field effect. However, recent theoretical and experimental studies have demonstrated that even in the absence of SOC, electrons can acquire orbital angular momentum under an external electric field[81, 177-

[181]. These discoveries give rise to orbitronics, an emerging research field that explores the utilization of orbital angular momentum for the manipulation of magnetic states[81]. Analogous to the SHE, the application of an external electric field $E$ induces an orbital current $J_L$ in a direction perpendicular to $E$, a phenomenon known as the orbital Hall effect (OHE)[177-179]. In addition to the OHE, the orbital Rashba-Edelstein effect (OREE) can generate orbital currents at the interface of NM/FM heterostructures, similar to the conventional REE[182-184]. Unlike spin angular momentum, orbital angular momentum does not directly exchange-couple with magnetic moments. Consequently, within an FM layer, the $J_L$ is first converted into a spin current $J_S$ via SOC (Fig. 7(a)), which subsequently exerts OTs on the magnetization[79, 82, 185-187]. In NM/FM heterostructures, OT and conventional SOT coexist, but their contributions can differ in magnitude and sign. Specifically, when the NM exhibits negative spin-orbit correlation ($R_{NM}<0$), as in Ta[79], W[80], Zr[87, 188], V[83, 189], Cr[82], and Mn[190], the $J_S$ generated in the NM is opposite to the $J_L$. If the $L$-$S$ conversion coefficient ($\eta_{L\text{-}S}$) of the FM (or the orbital-spin conversion layer) is positive, as in Pt[82], Co[82, 84, 191], CoFe[79], and Ni[79, 82], the $J_S$ converted from the $J_L$ is opposite to the $J_S$ from the SHE or REE. Conversely, if $R_{NM} > 0$ (e.g., Ru[85], Pt[192]) with $\eta_{L\text{-}S}>0$, or $R_{NM} < 0$ with $\eta_{L\text{-}S}<0$ (e.g., Ta[179], W[179], and Gd[82]), the $J_S$ converted from the $J_L$ has the same sign as the $J_S$ from the SHE or REE. As a result, the net torque is determined by the interplay between OT and SOT, which can be expressed as: $\xi_{\text{SOT,total}} = \xi_{\text{OT}} + \xi_{\text{SOT}}$, where $\xi_{\text{SOT}}$ is primarily dependent on the $\theta_{\text{SH}}$ of the NM, whereas OT efficiency ($\xi_{\text{OT}}$) depends on both the orbital Hall angles ($\theta_{\text{OH}}$) of NM and $\eta_{L\text{-}S}$ in the FM. The effective $\theta_{\text{OH}}$ is defined as: $\theta_{\text{OH}}^{\text{eff}} = \theta_{\text{OH}} \cdot \eta_{L\text{-}S} = (2e/\hbar)\sigma_{\text{OHE}}\eta_{L\text{-}S}/\sigma_j$, where $\sigma_{\text{OHE}}$ denotes the OHC. Therefore, selecting NM materials with high $\sigma_{\text{OHE}}$ and FM layers with a high $\eta_{L\text{-}S}$ are essential for maximizing $\theta_{\text{OH}}^{\text{eff}}$. To enhance the $\theta_{\text{OH}}^{\text{eff}}$, an insert layer (e.g. Pt, Gd, Tb, etc.) with strong SOC has been proposed to efficiently convert the $J_L$ into the $J_S$ with high $\eta_{L\text{-}S}$ (Fig. 7(b))[25, 82, 84], and more details will be discussed later. Moreover, the orbital diffusion length ($\lambda_{\text{OH}}$) can be obtained from the equation[81]: $\xi_{\text{DL}} = \theta_{\text{OHE}}^{\text{eff}}\left(1 - \text{sech}\left(\frac{t_{\text{OT}}}{\lambda_{\text{OH}}}\right)\right) + \xi_0$, where $t_{\text{OT}}$ is the thickness of orbital source material, $\xi_{\text{DL}}$ is anti-damping torque efficiency, and $\xi_0$ is a constant arising from interfacial origins, such as the OREE and interfacial SOT. $\lambda_{\text{OH}}$ is much longer than the spin diffusion length, which has been used to experimentally separate the OT contribution from SOT[191]. The long orbital diffusion length, broad material selection, and high OHC make OT highly efficient for magnetization switching. These attributes offer promising pathways for the development of low-power, high-stability orbitronic memory and logic devices. It is worth noting that the backflow effect may diminish the advantage of the long orbital diffusion length. When $t_{\text{OT}}$ is smaller than $\lambda_{\text{OH}}$, the orbital current is bounced back at the NM/FM interface and neutralizes with other orbital currents. Conversely, if $t_{\text{OT}}$ significantly exceeds $\lambda_{\text{OH}}$, the backflow effect can be effectively suppressed, but this will lead to increased current shunting and higher energy consumption. Therefore, a comprehensive consideration is needed to select the appropriate thickness of orbital source material in the design of OT devices.

## 3.2. Materials for orbital Hall effect

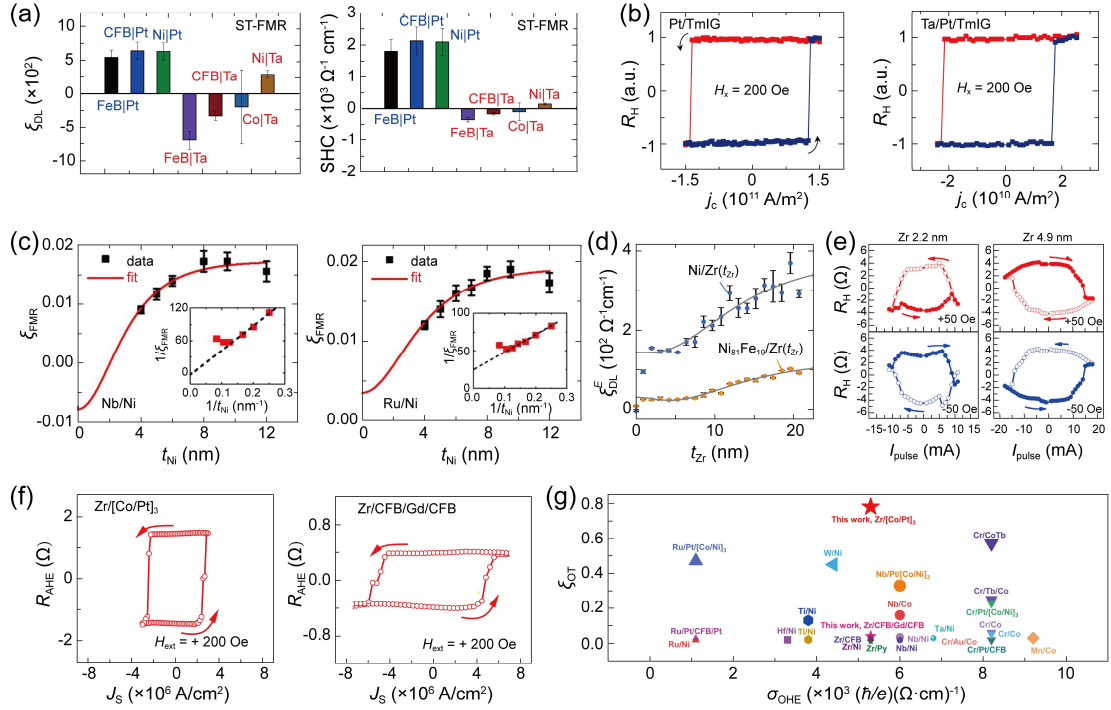

**Fig. 8.** The orbital Hall effect in 4$d$ and 5$d$ metals and the magnetization switching driven by them. (a) The anti-damping torque efficiency ($\xi_{DL}$) and the effective spin Hall conductivity ($\sigma_{DL}$) for various ferromagnet/non-magnet (FM/NM) bilayers (FM = FeB, CoFeB, Co, Ni, NM = Pt, Ta). Reprinted with permission from Ref. [79]. Copyright 2021 by Springer Nature. (b) Current-induced magnetization switching curves for the Ta/TmIG and Ta/Pt/TmIG. Reprinted with permission from Ref. [193]. Copyright 2023 by American Chemical Society. (c) The spin-torque ferromagnetic resonance efficiency of Nb/Ni and Ru/Ni heterostructure. Reprinted with permission from Ref. [86]. Copyright 2023 by American Physical Society. (d) The Zr-layer thickness ($t_{Zr}$) dependence of $\xi_{DL}$ for the Zr/Py (orange) and Zr/Ni (blue) bilayers. Reprinted with permission from Ref. [188]. Copyright 2023 by American Physical Society. (e) Current-induced magnetization switching for Zr (2.2 nm)/CoFeB/MgO and for Zr (4.9 nm)/CoFeB/MgO. Reprinted with permission from Ref. [194]. Copyright 2020 by American Chemical Society. (f) Current-induced orbital torque switching for Zr/[Co/Pt]$_3$ and Zr/CoFeB/Gd/CoFeB. (g) The summary of orbital torque efficiency ($\xi_{OT}$) as a function of orbital Hall conductivity ($\sigma_{OHE}$) for the orbital Hall materials. Reprinted with permission from Ref. [87]. Copyright 2024 by Springer Nature.

Since the OHE is independent of SOC, a wide range of materials have been theoretically predicted to exhibit strong OHC, including 3$d$, 4$d$, and 5$d$ transition metals, $sp$-metals, and transition metal dichalcogenides (TMDs)[84, 178, 179, 195-204]. A comprehensive review of OHC values for various materials, as well as $\theta_{OH}^{eff}$ and $\lambda_{OH}$ in different heterostructures, can be found in a recent review[81]. The SHE in heavy 5$d$ transition metals, such as Ta[14], W[99], and Pt[97], has been extensively studied and successfully implemented in SOT devices. However, theoretical and experimental studies indicate that these materials also exhibit an exceptionally large $\sigma_{OHE}$ on the order of $\sim 10^5$ ($\hbar/2e$) ($\Omega \cdot$m)$^{-1}$, surpassing their SHC[79, 178]. Lee et al. investigated the anti-damping torque efficiency ($\xi_{DL}$) in Pt/CoFeB, Ta/CoFeB, and Pt/Ni bilayers using ST-FMR[79]. They found that $\xi_{DL}$ follows the sign of SHC in Pt/CoFeB and Ta/CoFeB, consistent with conventional SOT mechanisms. However, an anomalous behavior is observed in Ta/Ni bilayers, where the sign of the net torque is opposite to the SHC of Ta[79] (Fig. 8(a)). This phenomenon is well explained by the superposition of $\sigma_{SHE}^{Ta}$ and $\eta_{L-S}^{Ni}\sigma_{OHE}^{Ta}$. Specifically, when $\sigma_{SHE}^{Ta} > 0$ and the contribution from OHE surpasses that of SHE, the resulting net torque changes sign, providing direct evidence of OT. Further experimental validation of OHE-driven torque was provided by Li et al., who demonstrated that in a Ta/Pt/thulium iron garnet ($Tm_3Fe_5O_{12}$, TmIG) heterostructure, the orbital current generated in Ta can be converted into a spin current in Pt, thereby enhancing the torque for magnetization switching (Fig. 8(b))[193]. This finding challenges the conventional belief that stacking materials with opposite $\theta_{SH}$ (e.g., Ta and Pt) will necessarily lead to a reduction in net torque efficiency. Moreover, OT efficiency is highly dependent on the choice of FMs. In W/Ni and Pt/Ni bilayers, the torque efficiency increases with the Ni thickness ($t_{Ni}$), even when $t_{Ni}$ exceeds 10 nm. This behavior is in stark contrast to conventional SOT and is attributed to the significantly longer diffusion length of orbital currents compared to spin currents[80, 192]. Moreover, in Hf/Ni bilayers, the crystallinity of the Ni layer plays a crucial role in determining OT efficiency, a phenomenon that is not observed in SOT driven by SHE[205]. Theoretical calculations have also predicted significant OHC in 4$d$ transition metals[178]. Experimental investigations using ST-FMR have confirmed these predictions. In Nb/Ni and Ru/Ni bilayers, $\theta_{OHE}^{eff}$ is measured to be 0.018 and 0.019, respectively, with $\lambda_{OH}$ of approximately 2.2 nm and 2.7 nm[86] (Fig. 8(c)). In Zr/Ni bilayers, the $\xi_{DL}$ increases as a function of both $t_{Ni}$ and $t_{Zr}$[188] (Fig. 8(d)). Notably, a sign change in $\xi_{DL}$ is observed[194], similar to the behavior reported in Ta/Ni bilayers[79]. This sign reversal indicates a transition from SHE-dominated to OHE-dominated torques, highlighting the dominant role of orbital currents in magnetization dynamics. Furthermore, current-induced magnetization switching is demonstrated in Zr/CoFeB/MgO heterostructures (Fig. 8(e)). Samples with 2.2 nm and 4.9 nm Zr exhibit opposite switching behavior, providing direct evidence of a crossover from conventional SHE-driven SOT to an OHE-driven mechanism[194]. Recently, Yang et al. achieved perpendicular magnetization switching in [Co/Pt]$_3$ and CoFeB/Gd/CoFeB heterostructures using Zr as an orbital current source[87]. The Zr/[Co/Pt]$_3$ sample exhibits the highest reported $\xi_{OT}$ (~0.78) to date, attributed to the exceptionally large $\sigma_{OHE}$ of Zr (~$5.3 \times 10^5$ ($\hbar/2e$) ($\Omega \cdot m$)$^{-1}$)[188] and the large $\eta_{L-S}$ of [Co/Pt]$_3$ (Fig. 8(f))[87]. Figure 8(g) summarizes the $\xi_{OT}$ values as a function of $\sigma_{OHE}$ for various orbital Hall materials, providing a comparative perspective on the efficiency of OT generation[87].

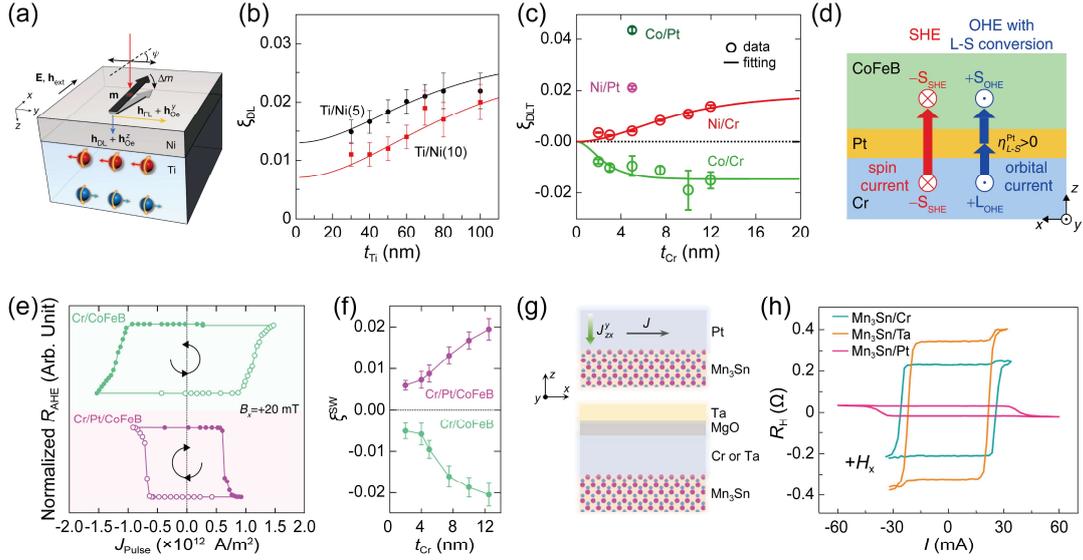

**Fig. 9.** The orbital Hall effect in 3*d* metals and the magnetization switching driven by them. (a) Schematic illustration of orbital torque in the Ti/Ni bilayer. (b) Ti thickness ($t_{Ti}$) dependence of $\xi_{DL}$ of the Ti/Ni bilayers. Reprinted with permission from Ref. [26]. Copyright 2023 by Springer Nature. (c) Cr thickness ($t_{Cr}$) dependent $\xi_{DL}$ of Ni/Cr (red) and Co/Cr (light-green) samples. (d) Schematic illustration of the orbital-to-spin conversion in the Cr/Pt/CoFeB heterostructure. The spin (orbital) angular momentum is represented by *S* (*L*). The subscript OHE or SHE marks the source of *S* (*L*). (e) Magnetization switching curves of Cr/CoFeB and Cr/Pt/CoFeB samples under an assisting magnetic field ($B_x$) of +20 mT. The green and magenta symbols represent the samples without and with the Pt insertion layer, respectively. (f) $t_{Cr}$-dependent switching efficiency of the Cr/CoFeB and Cr/Pt/CoFeB samples. Reprinted with permission from Ref. [82]. Copyright 2021 by Springer Nature. (g) Schematic of Mn$_3$Sn/X with X = Cr, Ta, or Pt structures. (h) Current dependence of Hall resistance at room temperature for Mn$_3$Sn/X with X = Cr, Ta, or Pt, with a positive in-plane assistant field. Reprinted with permission from Ref. [206]. Copyright 2023 by American Chemical Society.

3*d* transition metals have been overlooked in conventional SOT due to their weak SOC. However, recent research has highlighted their strong OHC, cost-effectiveness, environmental friendliness, and abundance, making them promising candidates for OT applications[195]. In Ti/Ni bilayers, OT has been experimentally demonstrated via magneto-optic Kerr effect (MOKE), yielding $\theta_{OHE}^{eff}$ = 0.021 and $\lambda_{OH}$ = 60 nm (Figs. 9(a) and 9(b))[26]. Additionally, Hayashi et al. obtained $\theta_{OHE}^{eff}$ = 0.13 and $\lambda_{OH}$ = 47 nm for Ti/Ni bilayers using ST-FMR measurements[80]. Similarly, Cr has been shown to exhibit $\theta_{OHE}^{eff}$ = 0.28 and $\lambda_{OH}$ = 6.6 nm, as determined via MOKE[207]. Interestingly, Cr-based bilayers exhibit a sign reversal in OT depending on the adjacent FM. Lee et al. found that Cr/Co bilayers generate torque efficiency with an opposite sign compared to Cr/Ni, Pt/Co, and Pt/Ni bilayers (Fig. 9(c))[82]. This behavior can be attributed to Cr's large $\sigma_{OHE}^{Cr}$ (8.2×10$^5$ ($\hbar$/2e) ($\Omega$·m)$^{-1}$) and its negative $\sigma_{SHE}^{Cr}$ (−130 ($\hbar$/2e) ($\Omega$·m)$^{-1}$), which have opposite signs[195]. In Cr/Co bilayers, the $\eta_{L-S}^{Co}$ of Co is close to zero, leading to a dominant SOT from the SHE and a negligible $\xi_{OT}$. Conversely, in Cr/Ni and Cr/Pt bilayers, where $\eta_{L-S}^{Ni} > 0$ and $\eta_{L-S}^{Pt} > 0$, the OHE dominates, resulting in a positive $\xi_{OT}$. As shown in Fig. 9(d), in Cr/Pt/CoFeB heterostructures with PMA, the $J_S$ converted from the $J_L$ follows the relation $\sigma_{OHE}^{Cr}\eta_{L-S}^{Pt} > 0$, whereas the spin current from the SHE follows $\sigma_{SHE}^{Cr} < 0$. This results in an anticlockwise switching curve with positive $\theta_{OHE}^{eff}$. Conversely, the control Cr/CoFeB bilayer exhibits a counterclockwise switching curve with negative $\theta_{OHE}^{eff}$ (Fig. 9(e)) [82]. Notably, the dependence of switching efficiency on Cr thickness is similar in both structures, suggesting that the torque primarily originates from Cr rather than Pt (Fig. 9(f))[82]. Beyond FMs, OT has also been utilized to manipulate AFMs. Xie et al. demonstrated AFM switching of Mn$_3$Sn layers using OT from Cr[206]. Cr/Mn$_3$Sn sample exhibits the same switching polarity as Mn$_3$Sn/Ta, consistent with the positive $\theta_{OH}^{eff}$ of Cr (Figs. 9(g) and (h)). Similar AFM switching behavior has also been reported using OT from Mn[208].

First-principles calculations predict that certain $sp$ metals Al[202] and XIV group elemental crystal (Si, Ge, and $\alpha$-Sn)[209] exhibit negative $\sigma_{OHE}$. Moreover, FnMn alloys have been reported to possess an exceptionally high $\sigma_{OHE}$ of 5.45×10$^7$ ($\hbar/2e$) ($\Omega \cdot$m)$^{-1}$, enabling magnetization switching in PMA CoPt heterostructures[210]. For 2D materials, OHC studies have largely focused on theoretical predictions. Tokatly revealed that in $p$-doped graphene, the OHE originates from Berry phase accumulation as heavy and light holes move around the degeneracy point at the Brillouin zone center[211]. Bhowal further clarified that the OHC in gapped graphene is distinct from the previously reported valley Hall effect[212]. In TMDs such as MX$_2$ (M = Mo, W; X = S, Se, Te; NbS$_2$), the reported $\sigma_{OHE}$ (~10$^5$-10$^7$ ($\hbar/2e$) ($\Omega \cdot$m)$^{-1}$) is significantly larger than $\sigma_{SHE}$ and exhibits an opposite sign, indicating the potential of TMDs in orbitronic devices[196, 198, 200, 201, 213]. Furthermore, MX$_2$ (M = Pt, Pd; and X = S, Se, and Te) exhibits an intrinsic OHC (~ 10$^5$ ($\hbar/2e$) ($\Omega \cdot$m)$^{-1}$) due to orbital texture effects around momentum-space valleys[204]. Liu et al. proposed that the OHE in graphene, TMDs, and topological AFMs is primarily driven by extrinsic mechanisms, suggesting that OT in these systems could be tuned or enhanced via doping[203]. Additionally, 2D FMs and AFMs have been actively studied. Chen et al. proposed that topological phase transitions can modulate the OHE in topological AFM MnBi$_2$Te$_4$[214]. Zhang et al. experimentally demonstrated magnetization switching in van der Waals FM Fe$_3$GaTe$_2$ driven by OT from Ti, achieving a switching current density of 1.6×10$^6$ A·cm$^{-2}$[215]. While theoretical predictions for OHE in 2D materials are promising, further experimental validation is needed to fully realize their potential in orbitronic devices.

### 3.3. Materials for orbital Rashba-Edelstein effect

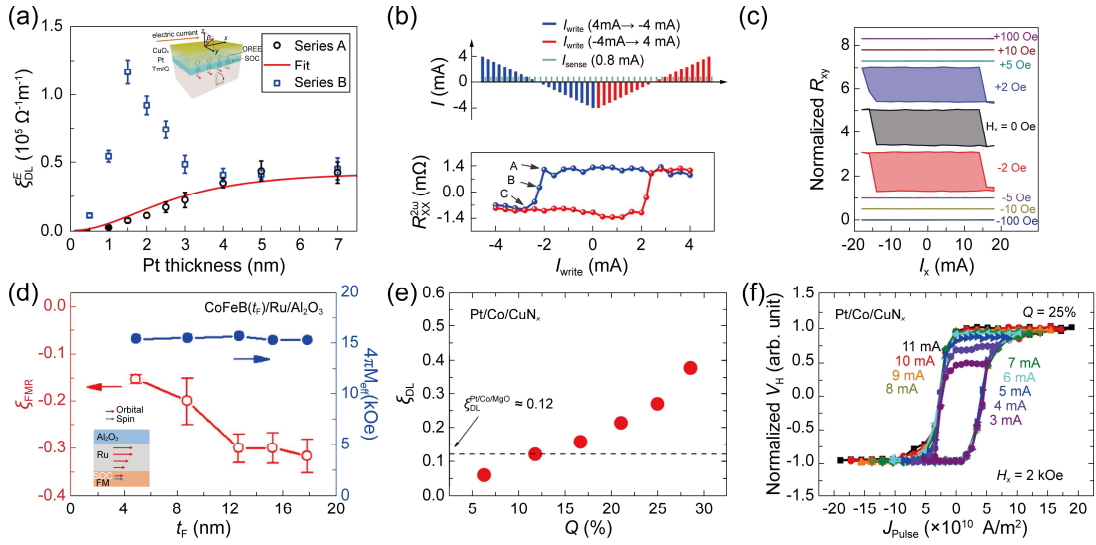

**Fig. 10.** The orbital torque efficiency of orbital Rashba-Edelstein effect material and the magnetization switching driven by orbital Rashba-Edelstein effect. (a) $\xi_{DL}$ as a function of Pt thickness for TmIG/Pt (series A) and TmIG/Pt/CuO$_x$ (series B). Reprinted with permission from Ref. [25]. Copyright 2020 by American Physical Society. (b) $R_{xx}^{2\omega}$-$I_{write}$ loop and its procedure for measuring. Reprinted with permission from Ref. [216]. Copyright 2023 by American Chemical Society. (c) Magnetization switching curves of TmIG (10 nm)/CoFeB (1.5 nm)/Cu (3 nm)/SiO$_2$ under a series of external fields ($H_x$) along the $x$-axis. Reprinted with permission from Ref. [217]. Copyright 2023 by Wiley-VCH GmbH. (d) Ferromagnetic thickness ($t_F$) dependence of torque efficiency of CoFeB/Ru/Al$_2$O$_3$. Reprinted with permission from Ref. [85]. Copyright 2022 by American Physical Society. (e) $\xi_{DL}$ as functions of nitrogen doping concentration ($Q$). (f) Switching loops with various maximum applied current densities. Reprinted with permission from Ref. [218]. Copyright 2022 by American Physical Society.

The OREE arises at the interfaces and surfaces of NM/FM heterostructures. In 2016, Ando's group first reported a significant spin torque in Cu/Py bilayers, generated by the light metal Cu, despite its weak SOC[219]. They attributed the robust anti-damping torque in CuO$_x$/Py to Berry curvature effects[220] and further demonstrated that oxygen-induced modifications of interfacial orbital hybridization could enhance the torque efficiency[221]. In 2020, Ding et al. observed a substantial enhancement in torque efficiency in TmIG/Pt heterostructures upon capping with a CuO$_x$ layer. They suggested that this enhancement originates from the orbital current generated at the CuO$_x$/Pt interface[25]. As illustrated in Fig. 10(a), the $\xi_{DL}$ in TmIG/Pt (2 nm)/CuO$_x$ (~0.01) is an order of magnitude higher than that in TmIG/Pt, attributed to the OREE-induced OT. Similar large torque efficiencies have been observed in FM/Cu/AlO$_x$[180, 184] and FM/Cu/CuO$_x$[187] trilayers, where the interfacial inversion symmetry breaking results in an orbital asymmetry that induces a Rashba-type orbital angular momentum texture[195]. Consequently, under an external electric field, a finite orbital angular momentum is generated, despite its equilibrium net value remaining zero[222, 223]. As $t_{Pt}$ increases, the $\xi_{DL}$ decreases, eventually reaching the same value as in TmIG/Pt when $t_{Pt}$ exceeds 4 nm, due to spin current dephasing in Pt (Fig. 10(a))[25]. Here, Pt serves as the conversion layer that transforms the $J_L$ into the $J_S$, while $CuO_x$ does not absorb $J_S$[224], effectively improving the $\xi_{DL}$. Harnessing the OREE, Huang et al. achieved field-free in-plane magnetization switching in CoFeB/$CuO_x$ devices without heavy metals (Fig. 10(b))[216]. Similarly, Xiao et al. demonstrated perpendicular magnetization switching in Pt/Co/Cu-$CuO_x$ heterostructures, facilitated by the combined effect of OT from the Cu/$CuO_x$ interface and SOT from Pt[225]. Furthermore, field-free perpendicular magnetization switching driven by OT has been realized in TmIG/CoFeB/Cu/$SiO_2$ heterostructures (Fig. 10(c))[217]. In this system, TmIG exhibits PMA, while CoFeB exhibits IMA. The interfacial exchange coupling induces a tilt in the magnetization directions of TmIG and CoFeB, leading to a strong dependence of the TmIG up/down states on the CoFeB magnetization orientation. Consequently, the $J_L$ generated in Cu is converted into the $J_S$ in CoFeB, exerting OTs on TmIG and enabling deterministic switching. Beyond the $CuO_x$-based system, OREE-induced OT has also been observed at Ru/$Al_2O_3$[85] and Co/$CuN_x$[218] interfaces. As shown in Fig. 10(d), the $\xi_{OT}$ of CoFeB(12 nm)/Ru/$Al_2O_3$ increases with CoFeB thickness and saturates at ~0.3 for $t_{CoFeB}$ = 12 nm[85]. Meanwhile, in Pt/Co/$CuN_x$ heterostructures, $\xi_{DL}$ varies between 0.06 and 0.4, depending on the nitrogen doping concentration in $CuN_x$ (Fig. 10(e)). This OT effect is further evidenced by magnetization switching with a critical current density of $5\times10^6$ A·cm$^{-2}$ (Fig. 10(f)).

## 4. Magnon torque
### 4.1. Mechanism of magnon torque

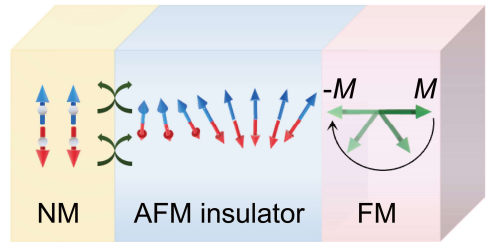

**Fig. 11.** The principal diagram of magnon torque.

A magnon is a quasiparticle that represents the quantized excitations of spin waves in magnetic materials[226, 227]. When magnons propagate collectively from one region to another, their coordinated motion gives rise to a magnon current. Unlike electron-mediated spin currents, magnon currents carry spin angular momentum without involving charge motion, making them particularly advantageous in magnetic insulators. This charge-free nature of magnon transport eliminates Joule heating caused by charge motion, a major limitation in conventional electron-mediated systems, thereby enabling energy-efficient spin information transport. Moreover, in magnetic insulators, magnons can propagate over micrometer-scale distances[88-90] at ultrafast velocities[91], further enhancing their potential for spin information processing. Finally, magnons exhibit excellent controllability and can be precisely manipulated using well-established techniques, including ultrafast thermal gradients[228, 229], attosecond laser pulses[230], microwave excitations[231], and electric fields[28, 232, 233]. Benefiting from the above attributes, magnons have emerged as a tool for the manipulation of magnetic moments, which manifests as domain wall motion and magnetization switching[234-236].

In a typical magnon device, the structure consists of an NM layer, an AFM insulator, and an FM, as illustrated in Fig. 11. When a charge current passes through the NM layer, it generates a spin current, leading to nonequilibrium spin accumulation at the NM/AFM interface. Through interfacial exchange interactions, the nonequilibrium electron spin current is converted into a magnon current within the AFM. This magnon current subsequently propagates through the AFM layer and exerts MTs on the adjacent FM, ultimately driving its magnetization switching. The NM layer in such devices is typically required to exhibit high charge-to-spin conversion efficiency and can be selected from the materials discussed above. The AFM insulating layer is commonly composed of NiO, which possesses a high Néel temperature, or the multiferroic material BiFeO$_3$. In the following sections, we will review MTs, categorizing the discussion based on different AFM insulating layers.

### 4.2. NiO-based magnon torque

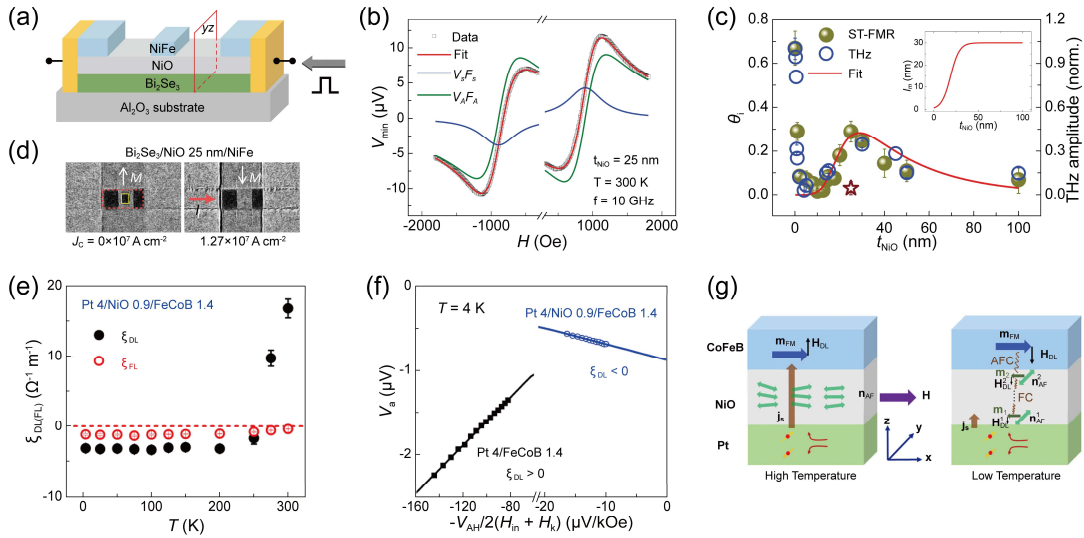

**Fig. 12.** Magnetization switching by magnon torques through an antiferromagnetic insulator NiO. (a) Illustration of the structure of magnon torques switching device with an isolated NiFe rectangle defined on top of the NiO layer. (b) A typical spin-torque ferromagnetic resonance signal from a Bi$_2$Se$_3$/NiO (25 nm)/Py (6 nm) device at 10 GHz and 300 K with fits. (c) The spin torque ratio ($\theta_i$) deduced from the spin-torque ferromagnetic resonance data and the terahertz emission amplitude as a function of the thickness of NiO ($t_{NiO}$) at 300 K. The star symbol marks the $\theta_i$ from the control device. (d) Magneto-optic Kerr effect images for magnon torque-driven magnetization switching in the Bi$_2$Se$_3$/NiO/Py device by injecting a pulsed current (*I*) along the +*x* axis. Reprinted with permission from Ref. [24]. Copyright 2019 by American Association for the Advancement of Science. (e) Temperature dependence of the anti-damping ($\xi_{DL}$) and field-like ($\xi_{FL}$) torque efficiencies per applied electric field in Pt (4 nm)/NiO (0.9 nm)/ CoFeB (1.4 nm). (f) Linear dependence of $V_a$ on $-V_{AH}/2(H_{in}+H_k)$ for Pt (4 nm)/NiO (0.9 nm)/ CoFeB (1.4 nm) and Pt (4 nm)/ CoFeB (1.4 nm) at 4 K with the slopes being anti-damping spin-orbit effective field $H_{DL}$, where $V_{AH}$, $H_{in}$, and $H_k$ are anomalous Hall voltage, in-plane magnetic field, and perpendicular anisotropy field, respectively. Reprinted with permission from Ref. [237]. Copyright 2021 by American Physical Society. (g) The schematic of the anti-damping spin-orbit effective fields ($H_{DL}$) for the Pt/NiO/CoFeB. Reprinted

with permission from Ref. [238]. Copyright 2022 by American Physical Society.

In 2015, the angular momentum transfer in AFM insulator NiO was demonstrated by ST-FMR in Pt/NiO/FeNi structure[239]. Later, Hyunsoo Yang's group achieved the in-plane magnetization switching utilizing MTs in a $Bi_2Se_3$/NiO/Py sandwich structure (Fig. 12(a))[24]. ST-FMR measurements (Fig. 12(b)) confirm the presence of in-plane anti-damping MT. Similar to the spin Hall angle in conventional SOT, the spin torque ratio is defined as $\theta_t = J_i/J_C$, whose dependence on the thickness of NiO ($t_{NiO}$) is shown in Fig. 12(c). When the NiO layer inserted between the $Bi_2Se_3$ and Py layers is very thin ($t_{NiO} < 1$ nm), the AFM ordering is absent at room temperature so that MT is negligible, and $\theta_t$ purely originates from electron-mediated torques. As $t_{NiO}$ increases to 2 nm, $\theta_t$ drops from 0.67 to 0 because electron tunneling is unallowed in an NM NiO layer. $\theta_t$ increases as the $t_{NiO}$ increases above 10 nm and arrives at the peak value (~0.3) at $t_{NiO} = 25$ nm, and then decreases. $\theta_t$ of the control sample, where a 6 nm MgO layer is inserted between $Bi_2Se_3$ and NiO, is negligible, excluding the possibility that observed torque originated from the NiO (25 nm)/Py interface. In current-induced switching measurements, the Py layer is patterned into a rectangular island to avoid the current shunting. The MOKE images demonstrate the magnetization switching of the FM with a critical switching current density ($J_C$) of ~$1.27 \times 10^7$ A·m$^{-2}$ induced by the in-plane anti-damping MT (Fig. 12(d)). This work is a milestone in MT. Afterwards, Zheng et al. fabricated an all-oxide heterostructure $SrRuO_3$/NiO/$SrIrO_3$ with PMA, in which the $J_C \sim 8.1 \times 10^5$ A·m$^{-2}$ is one order smaller than the conventional metallic system[240, 241]. Heavy metal Pt has been used as a spin source in MT devices, such as Pt/NiO/$Y_3Fe_5O_{12}$[242] and Pt/NiO/CoTb[243], which can enhance the torque efficiency in < 2 nm NiO-based devices. An interesting phenomenon in Pt/NiO/CoFeB system observed by Zhu et al is the sign reversal of SOT upon cooling down[237]. As shown in Figs. 12(e) and (f), as the temperature decreases, the anti-damping torque varies from positive to negative in Pt/NiO/CoFeB sample but remains positive in Pt/CoFeB sample. Following this work, Weishengzhao's group attributed the reversal to magnetic interactions associated with the exchange bias effect[238]. At low temperatures, the net moment at Pt/NiO interface ($\mathbf{m}_1$) receives a torque from the spin Hall current of Pt and is coupled to the net moment at NiO/CoFeB interface ($\mathbf{m}_2$), as shown in Fig. 12(g). Since $\mathbf{m}_2$ is exchange-coupled to CoFeB, its antiparallel alignment with CoFeB reverses the anti-damping torque compared to that at high temperatures. As the temperature increases above the blocking temperature of NiO, the magnetic moments in NiO are significantly reduced and the transfer of the torque through exchange coupling in NiO disappears[238]. Besides, the spin source in MT devices has been expanded to topological insulators. Shi et al.[244] developed the room-temperature MT structure $Bi_2Te_3$/NiO/CoFeB, which enables the switching of perpendicular magnetization with a $J_C$ of $4.1 \times 10^6$ A·cm$^{-2}$ and the MT efficiency is 0.33 with a magnon diffusion length of 26.6 nm. However, $Bi_2Te_3$ suffers from a high resistivity, leading to significant power consumption issues. Zhao et al. further optimized the power consumption of magnon devices by replacing $Bi_2Te_3$ with topological crystalline insulator SnTe, reducing power consumption to approximately 22 times lower than that of $Bi_2Te_3$[245].

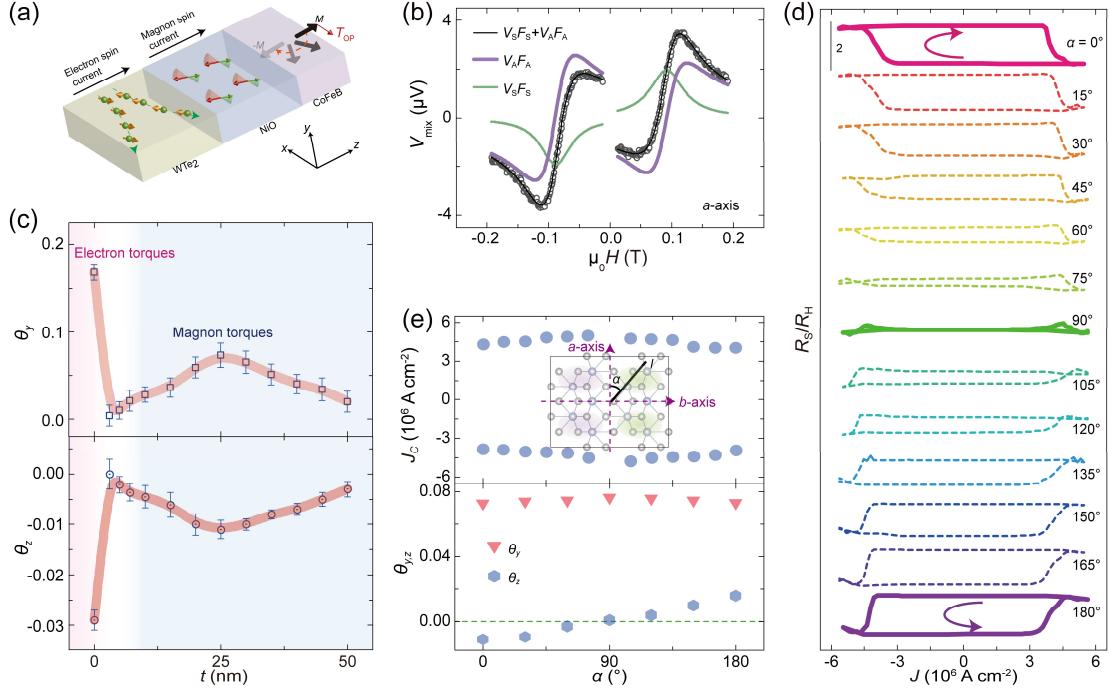

**Fig. 13.** Field-free switching of perpendicular magnetization by magnon torques. (a) Magnetization switching by magnon torque in WTe$_2$/NiO/CoFeB. (b) Spin-torque ferromagnetic resonance spectra of WTe$_2$ (8 nm)/NiO (25 nm)/Py with currents applied along the *a*-axis of WTe$_2$. (c) NiO thickness (*t*) dependence of in-plane torque efficiency ($\theta_y$) and out-of-plane torque efficiency ($\theta_z$) in WTe$_2$ (8 nm)/NiO (25 nm)/Py. (d) Current-induced field-free magnetization switching of the CoFeB layer. The switching is performed under different angles ($\alpha$) between the low-symmetry *a*-axis of WTe$_2$ and the current flow direction. (e) $\alpha$ dependence of critical switching current density ($J_C$), $\theta_y$, and $\theta_z$. Reprinted with permission from Ref. [246]. Copyright 2024 by Springer Nature.

In the above discussions, perpendicular magnetization switching relying on the in-plane anti-damping MT is found to be nondeterministic, necessitating an external magnetic field aligned with the current direction to achieve deterministic switching. The realization of field-free magnetization switching remains a key challenge for the practical application of magnon devices. Recently, Wang et al. successfully demonstrated field-free switching of perpendicular magnetization using out-of-plane anti-damping MTs[246]. They fabricated a WTe$_2$/NiO/CoFeB heterostructure, where the NM layer, WTe$_2$, breaks mirror symmetry along the low-symmetry *a*-axis. As illustrated in Fig. 13(a), when a charge current is applied along the *a*-axis, both *y*- and *z*-spins are generated. The electron-carried spins subsequently excite magnon currents in NiO, carrying *y*- and *z*-polarized magnon moments that exert in-plane and out-of-plane anti-damping MTs on CoFeB, respectively[246]. ST-FMR measurements with the current applied along the *a*-axis exhibit distinct differences in both the shape and amplitude of the mixed voltage ($V_{mix}$) under positive and negative magnetic fields (Fig. 13(b)), confirming the existence of the out-of-plane anti-damping MT. Moreover, they carried out $t_{NiO}$ dependence of in-plane ($\theta_y$) and out-of-plane ($\theta_z$) effective torque efficiencies, as shown in Fig. 13(c). The initial decrease in $\theta_y$ and $\theta_z$ for $t_{NiO}$ = 0-3 nm is attributed to the suppression of electron-mediated torques. As $t_{NiO}$ increases, both $\theta_y$ and $\theta_z$ rise, reaching peak values (~0.3) at $t_{NiO}$ = 25 nm, corresponding to the highest MT efficiencies. For $t_{NiO}$ > 25 nm, enhanced magnon-phonon and magnon-magnon interactions within NiO result in a reduction in MT efficiencies. To investigate the role of WTe$_2$'s crystal symmetry, magnetization switching is performed in the absence of an external magnetic field at varying angles ($\alpha$) between the low-symmetry a-axis of WTe$_2$ and the current direction (Fig. 13(d)). The extracted $J_C$, $\theta_y$, and $\theta_z$ are summarized in Fig. 13(e). A significantly higher $J_C$ is required when the current deviates from the a-axis, indicating that the observed out-of-plane anti-damping MT is closely related to the broken crystal symmetry of WTe$_2$. To reduce power consumption, Wang et al. fabricated PtTe$_2$/WTe$_2$/NiO (25 nm)/FM devices, achieving $\sigma_{s,y}$ of 1.5×10$^4$ ($\hbar/2e$) ($\Omega \cdot m$)$^{-1}$ and $\sigma_{s,z}$ of 2.6×10$^3$ ($\hbar/2e$) ($\Omega \cdot m$)$^{-1}$, resulting in a 190-fold reduction in power consumption compared to the Bi$_2$Te$_3$/NiO/CoFeB sample[246]. This work establishes an all-electric, low-power, perpendicular magnetization switching by MT, which is crucial for the advancement of magnon-based memory and logic devices.

### 4.3. BiFeO$_3$-based magnon torque

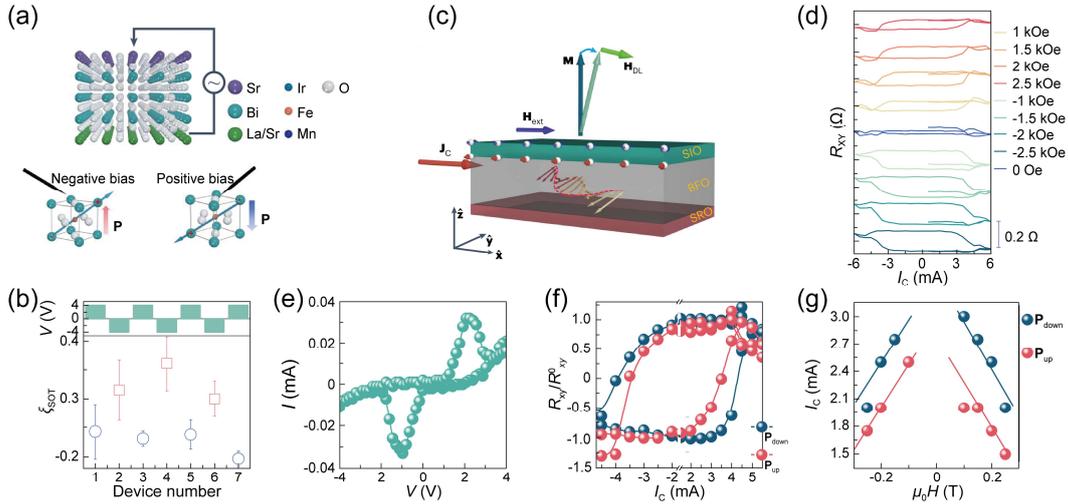

**Fig. 14.** The magnon torque in BiFeO$_3$ under the control of ferroelectric polarization. (a) Schematic of a La$_{0.7}$Sr$_{0.3}$MnO$_3$/BiFeO$_3$(BFO)/SrIrO$_3$(SIO) heterostructure under an external electric field, with the BiFeO$_3$ ferroelectric polarization under negative and positive biases. (b) The spin-orbit torque efficiency measured with the spin-torque ferromagnetic resonance technique as a function of bias voltages applied beforehand for La$_{0.7}$Sr$_{0.3}$MnO$_3$/BiFeO$_3$/SrIrO$_3$. (c) Schematic of SrRuO$_3$ (SRO) magnetization switching by the current-induced torques borne by the magnons transmitted through BiFeO$_3$. (d) The anomalous Hall resistance ($R_{xy}$) versus pulsed currents at 70 K under assisting magnetic fields with different polarities and strengths. (e) The current-voltage (I-V) curve during the ferroelectric switching experiments for the heterostructure SrRuO$_3$/BiFeO$_3$/SrIrO$_3$. (f) Ferroelectric polarization-controlled magnetization switching measurement. The magnetization switching behavior with the upward (downward) ferroelectric polarization of BiFeO$_3$ is shown in red (blue). $R_{xy}$ is normalized by $R_{xy}^0$ (anomalous Hall resistance at $I_C$ = 0). (g) The magnitudes of the critical switching current as a function of the assisting magnetic fields. The lines are a guide to the eye. Reprinted with permission from Ref. [28]. Copyright 2024 by Springer Nature.

When the AFM insulator NiO is replaced with a multiferroic material, an external electric field can modulate the Néel vector orientation of the multiferroic layer, thereby controlling the generation and propagation of magnon currents. BiFeO$_3$, a representative multiferroic material, exhibits coexisting antiferromagnetism and ferroelectricity, which are intrinsically coupled via magnetoelectric interactions[247, 248]. As illustrated in Fig. 14(a), the ferroelectric polarization of BiFeO$_3$ can be switched between an upward and downward state by applying a negative or positive bias exceeding the coercive voltage to the top SrIrO$_3$ layer[28]. To investigate the impact of ferroelectric polarization on magnon transport, La$_{0.7}$Sr$_{0.3}$MnO$_3$/BiFeO$_3$/SrIrO$_3$ heterostructures are characterized using ST-FMR[28]. The results indicate that the MT efficiency is significantly enhanced for upward polarization (negative bias) and reduced for downward polarization (positive bias), as shown in Fig. 14(b). This observation confirms that MTs in BiFeO$_3$ can be effectively regulated via ferroelectric polarization. Magnetization switching experiments are further conducted using SrRuO$_3$/BiFeO$_3$/SrIrO$_3$ heterostructures[28], as depicted in Figs. 14(c) and 14(d). The corresponding current-voltage (*I-V*) characteristics during ferroelectric switching are presented in Fig. 14(e), where current peaks are associated with the reversal of ferroelectric polarization. Magnetization switching measurements are performed under different BiFeO$_3$ polarization states, as shown in Fig. 14(f), revealing that the critical switching current is lower for upward polarization than for downward polarization (Fig. 14(g)). This trend corresponds to enhanced spin transmission, aligning well with the ST-FMR results. This study highlights the feasibility of using ferroelectric polarization as a voltage-controlled mechanism for tuning MTs. Subsequently, Chai et al. magnon-based logic devices based on BiFeO$_3$ systems[233].

## 5. Conclusion and perspectives

SOT has been rapidly advancing to meet the growing demands for low-power and field-free magnetization switching, offering promising prospects for applications in magnetic data storage, logic circuits, and high-frequency spintronic devices. We have reviewed recent advancements in electron- and magnon-mediated torques, with a particular focus on their generation mechanisms and the associated spin source materials. These materials include conventional metals, TIs, 2D materials, NCAFM, and altermagnets. Despite these significant advancements, several critical challenges remain in achieving highly efficient electron- and magnon-mediated SOT switching and elucidating the underlying physical mechanisms:

(1) The exploration of more efficient low-symmetry materials is crucial for advancing SOT devices. For instance, recent experiments observed the nonlinear Hall effect in NbIrTe$_4$, indicating lateral-symmetry breaking and that NbIrTe$_4$ could serve as a potential source of *z*-spin currents[249]. However, its application in perpendicular magnetization switching has yet to be realized. Additionally, RuO$_2$ remains the only altermagnet experimentally verified to generate *z*-spin currents through the SSE. Expanding the range of altermagnets is essential, including strategies such as inducing a phase transition from conventional AFMs to altermagnets via strain engineering[250]. Notably, the epitaxial strain may also enable *g*-wave altermagnets such as CrSb and MnTe to generate *z*-polarized spin currents[251]. Moreover, the generation and utilization of $z$-spin currents in SOT devices have yet to be demonstrated in the $d$-wave altermagnet $Mn_5Si_3$, highlighting the need for further experimental and theoretical investigations.

(2) Although the spin-split electronic bands and their associated phenomena in $RuO_2$ have been extensively studied and widely utilized, the exact magnetic ground state of $RuO_2$ remains controversial[252-255]. This ongoing debate may stem from variations in strain conditions[256, 257] and fluctuations in the Ru/O stoichiometric ratio[252] across different $RuO_2$ samples. Therefore, further comprehensive experimental investigations are urgently required to unambiguously determine the antiferromagnetic properties of $RuO_2$.

(3) While significant progress has been made in the study of orbital currents, further exploration is required to fully understand the underlying transport mechanisms, develop direct measurement techniques, and optimize OT performance. Many theoretically predicted orbital materials, such as graphene[211, 212] and 2D TMDs, necessitate and are undergoing experimental validation. For example, the orbital texture has been experimentally observed in 2H-$WSe_2$[258] and $TiSe_2$[259]. Concurrently, the discovery of novel materials is critical, particularly ferromagnets with high $\eta_{L-S}$ and orbital source materials with large OHC, which are expected to enhance the practical applicability of OTs. Furthermore, $z$-polarized orbital currents have been demonstrated in $IrO_2$[260], which potentially enables field-free magnetization switching through OTs.

(4) OT concept and orbital diffusion length remain controversial. For instance, Liu et al. proposed that the positive anti-damping torque efficiency of Ta/Ni system is a misinterpretation of a non-negligible self-induced ST-FMR signal of the FM layer, rather than OT[261]. Using first-principles scattering calculations, Rang et al. demonstrated that orbital angular momentum is relaxed within 1-2 atomic scales in transition metal/FM bilayers, challenging claims of nonlocal orbital current[262]. Additionally, Song et al., using the first-principles calculations, revealed that OHE contributions in transition metal/FM bilayers are often overshadowed by SHE or canceled by self-torques in ferromagnets like Ni[263]. These debates highlight the need for unified methodologies to distinguish OT and SOT, especially given the short orbital relaxation lengths predicted by first-principles calculations.

(5) In MT devices, NM layers are predominantly composed of conventional heavy metals (Pt) or 2D materials ($Bi_2Se_3$, $Bi_2Te_3$, and $WTe_2$) that serve as spin current sources. Expanding these materials to light metals with high OHC could significantly enhance the efficiency of magnon devices. Moreover, current research on AFM insulating layers has primarily focused on NiO and $BiFeO_3$. Replacing these with 2D AFM insulators may enable more efficient magnon conversion due to the smooth interfaces inherent to 2D materials. This advancement could ultimately lead to the development of MT devices entirely composed of 2D materials, further optimizing their performance and scalability.

(6) Integrating low-symmetry materials into SOT-MRAMs presents several technical challenges. The field-free switching strategy with low-symmetry materials relies on specific crystal and spin symmetries, which are often disrupted in the polycrystalline films commonly employed in commercial spintronic devices. Moreover, these materials (particularly 2D materials) face issues of large-scale growth and thermal stability during MgO-MTJs integration and complementary metal-oxide-semiconductor (CMOS) fabrication. To overcome these limitations, it is crucial to develop spin source materials that can keep thermal stability above 350 °C and generate $z$-polarized spins independently of crystal or magnetic symmetry, thereby enhancing their practical applicability.

**Acknowledgment**

J. L., B. B., and Z. Y. contributed equally to this work. This work was supported by the National Natural Science Foundation of China (U24A6002, 12174237 (X. X.), 52471253 (F. W.), 12404091 (J. L.), 52171183 (Z. Q.)). X. X. acknowledges the support from National Key Research and Development Program of China (Grant nos. 2022YFB3505301). F. W. acknowledges the support from the Fund Program for the Scientific Activities of Selected Returned Overseas Professionals in Shanxi Province (20240019), and Central Government's Special Fund for Local Science and Technology Development (YDZJSX2024D058). J. L. acknowledges the support from the Basic Research Plan of Shanxi Province (202403021212016). Z. Y. acknowledges the support from the Fundamental Research Program of Shanxi Province (202403021222252) and the Higher Education Science and Technology Innovation Plan Project of Shanxi (2024L146).